\documentstyle[12pt]{article}
\begin{document}

\title{\sf General scalar interaction in the supersymmetric FRW model }

\author{\sc  V.I. Tkach$^a$ \thanks{E-mail: vladimir@ifug1.ugto.mx},
\and J.J. Rosales $^{a,b}$\thanks{E-mail: juan@ifug3.ugto.mx}, 
\and J. Socorro $^a$\thanks{E-mail: socorro@ifug4.ugto.mx} }
\date{}
\maketitle

\begin{center}
{$^a$Instituto de F\'{\i}sica de la Universidad de Guanajuato,\\
Apartado Postal E-143, C.P. 37150, Le\'on, Guanajuato, M\'exico\\
$^b$ Ingenier\'{\i}a en computaci\'on, Universidad del Baj\'{\i}o\\
Av. Universidad s/n Col. Lomas del Sol, Le\'on, Gto., M\'exico}
\end{center}

\maketitle
\begin{abstract}
In this work we have constructed the most general action for a set of 
complex homogeneous scalar supermultiplets interacting with the scale 
factor in the supersymmetric FRW model. It is shown, that local conformal 
time supersymmetry leads to the scalar fields potential, which is defined 
in the same combination: K\"ahler potential and superpotential as in 
supergravity (or effective superstring) theories. This scalar fields
potential depends on arbitrary parameter $\alpha$, which is not fixed
by conformal time supersymmetry. 
\end{abstract}

\newpage

The study of supersymmetric minisuperspace models have lead to 
important and interesting results. To find the physical states it is 
sufficient to solve the Lorentz and supersymmetric constraints$^{1-3}$. 
 Some of this results have already been presented in 
two comprehensive and organized works: a book$^4$ and an 
extended review$^5$. 

In the previous works$^{6,7}$ we have proposed a new approach to the 
study of supersymmetric quantum cosmology. The main idea is to extend 
the group of local time reparametrization of the cosmological models to 
the $n=2$ local conformal time supersymmetry. For this purpose the odd 
``time" parameters $\eta, \bar\eta$ were introduced (where $\bar\eta$ is 
the complex conjugate to $\eta$), which are the 
superpartners of the usual time parameter $t$. After we have introduced 
the new ``time" parameters our new functions, previously only time $t$ 
functions, become superfunctions depending on $(t, \eta, \bar\eta)$, which 
are called superfields. Following the superfield procedure 
we have constructed the superfield action for the cosmological models, 
which is invariant under the $n=2$ local conformal time supersymmetry. The 
fermionic superpartners of the scale factor and of the homogeneous scalar 
fields in the quantum level are elements of the Clifford algebra.

In the present work we have studied the supersymmetric FRW model 
interacting with a set of $n$ complex homogeneous scalar 
supermatter fields. We have shown, that in this case the potential of 
scalar matter fields is a function of the K\"ahler potential and one 
arbitrary parameter $\alpha$. Furthermore, when $\alpha = 1$ the 
scalar fields potential becomes energy vacuum of the scalar fields 
interacting with chiral matter multiplets in the case of $N =1$ 
supergravity theory$^8$. 
 
Let's begin by considering the FRW action
\begin{equation} 
S_{grav} = \frac{6}{ 8\pi G_N}\int \left(-\frac{R \, {\dot R}^2}{2N} + 
\frac{1}{2}k NR + \frac{d}{dt} \left(\frac{R^2 \dot R}{2N}\right) \right )dt, 
\label{action}
\end{equation}
for $k = 1,0,-1$ stands for a spherical, plane and hyperspherical 
three-space respectively, $\dot R = \frac{dR}{dt}$, $G_N$ is the Newtonian 
gravitational constant, $N(t)$ is the lapse function and $R(t)$ is the scale
factor depending only on $t$. In this work we will apply the system units 
$c=\hbar=1$. 

It is well known, that the action (\ref{action}) preserves the invariance 
under time reparame\-tri\-za\-tion,

\begin{equation}
t^\prime \to t + a(t), 
\label{reparame} 
\end{equation}
if $R(t)$ and $N(t)$ are transformed as

\begin{equation}
\delta R = a\dot R,\qquad\qquad \delta N = (aN\dot ) \,. 
\label{transforme}
\end{equation}   
 
In order to obtain the superfield formulation of the action (\ref{action}) 
in ref. [6] the transformations of the time reparametrization 
(\ref{reparame}) were extended to the $n=2$ local conformal time 
supersymmetry $(t, \eta, \bar\eta)$. These transformations 
can be written as

\begin{eqnarray} 
\delta t &=& {I\!\! L}(t,\eta,\bar \eta) + \frac{1}{2}\bar \eta D_{\bar\eta}
{I\!\!L}(t,\eta,\bar \eta) -\frac{1}{2}\eta D_{\eta} 
{I\!\!L}(t,\eta,\bar \eta),\nonumber\\
\delta \eta &=& \frac{i}{2}D_{\bar\eta} {I\!\!L}(t,\eta,\bar \eta), 
\label{superfunction}\\
\delta \bar \eta &=& -\frac{i}{2} D_{\eta} {I\!\!L}(t,\eta,\bar \eta)
\nonumber
\end{eqnarray}
with the superfunction ${I\!\!L}(t,\eta,\bar\eta)$ written as

\begin{equation}
{I\!\!L}(t,\eta,\bar\eta) = a(t) + i\eta \bar\beta^\prime(t) + i\bar\eta 
\beta^\prime(t) + b(t)\eta \bar\eta, 
\label{super}
\end{equation}
where $D_{\eta} = \frac{\partial}{\partial \eta} + i\bar\eta 
\frac{\partial}{\partial t}$ and 
$D_{\bar\eta} = -\frac{\partial}{\partial \bar\eta} - 
i\eta\frac{\partial}{\partial t}$ are the supercovariant derivatives of the 
global conformal supersymmetry, which have dimension $[D_\eta]=l^{-1/2}$,
$a(t)$ is a local time reparametrization parameter, 
$\beta^\prime(t) = N^{-1/2}\beta(t)$ is the Grassmann complex parameter
of the local conformal susy transformations (\ref{superfunction}) and 
$b(t)$ is the parameter of local $U(1)$ rotations on the complex $\eta$. 

The superfield generalization of the action (\ref{action}), which is 
invariant under the transformations (\ref{superfunction}), was found in our 
previous work$^6$ and has the form 
\begin{eqnarray} 
S_{grav} &=& \frac{6}{\kappa^2}\int \left \{ -\frac{{I\!\!N}^{-1}}{2}{I\!\!R} 
D_{\bar \eta}{I\!\!R} 
D_{\eta}{I\!\!R}+ \frac{\sqrt{k}}{2} {I\!\!R}^2  + 
\frac{1}{4}D_{\bar \eta}\left ({I\!\!N}^{-1}{I\!\!R}^2 
D_{\eta}{I\!\!R}\right ) \right. \label{superaction} \\
&&-\left. \frac{1}{4}D_{\eta}\left ({I\!\!N}^{-1}{I\!\!R}^2 
D_{\bar \eta} {I\!\!R}\right ) \right \} d\eta d\bar \eta dt, \nonumber
\end{eqnarray}
where we introduce the parameter $\kappa^2 = {8\pi G_N}$. We can see, that 
this action is hermitian for $k = 0,1$. The last two terms in 
(\ref{superaction}) are a total derivative, which are necessary when we 
consider interaction and 
${I\!\!N}(t, \eta, \bar\eta)$ is a real one-dimensional gravity superfield, 
which has the form

\begin{equation}
{I\!\!N}(t, \eta, \bar\eta) = N(t) + i\eta\bar\psi^\prime(t) + i\bar\eta 
\psi^\prime(t) + \eta \bar\eta V^\prime(t), 
\label{superfi}
\end{equation}
where $\psi^\prime(t)= N^{1/2}\psi(t)$, $\bar\psi^\prime(t)= 
N^{1/2}\bar\psi(t)$ and $V^\prime(t)= NV + \bar\psi \psi$. This superfield 
transforms as

\begin{equation}
\delta {I\!\!N}=  ({I\!\!L}{I\!\!N}\dot ) + \frac{i}{2}D_{\bar\eta}{I\!\!L}
D_{\eta}{I\!\!N} + \frac{i}{2}D_{\eta}{I\!\!L}D_{\bar\eta}{I\!\!N}\, . 
\label{coordinate}
\end{equation}
The components of the superfield 
${I\!\!N}(t,\eta,\bar\eta)$ in (\ref{superfi}) are gauge fields of the 
one-dimensional $n=2$ extended supergravity. 

The superfield ${I\!\!R}(t, \eta, \bar\eta)$ may be written as
\begin{equation}
{I\!\!R}(t, \eta, \bar\eta)= R(t) + i\eta\bar\lambda^\prime(t) + 
i\bar\eta \lambda^\prime(t) + \eta \bar\eta B^\prime(t), 
\label{super-R}
\end{equation}
where $\lambda^\prime(t)= \frac{\kappa N^{1/2}}{\sqrt{R}}\lambda(t),$ 
$\bar\lambda^\prime(t)= \frac{\kappa N^{1/2}}{\sqrt{R}}\bar\lambda(t)$ 
and $B^\prime(t)= \kappa NB - \frac{\kappa}{6\sqrt{R}}(\bar\psi 
\lambda - \psi \bar\lambda)$.
The transformation rule for the real scalar
superfield ${I\!\!R}(t, \eta, \bar\eta)$  is

\begin{equation}
\delta {I\!\!R}= {I\!\!L}\dot{I\!\!R}+ \frac{i}{2}D_{\bar\eta}{I\!\!L}
D_{\eta}{I\!\!R} + \frac{i}{2}D_{\eta}{I\!\!L}D_{\bar\eta}{I\!\!R}. 
\label{genera}
\end{equation}
The component $B(t)$ in (\ref{super-R}) is an auxiliary 
degree of freedom; $\lambda(t)$
and $\bar\lambda(t)$ are the fermionic superpartners of the scale factor
$R(t)$. The superfield transformations (\ref{coordinate},\ref{genera}) 
are the generalization 
of the transformations for $N(t)$ and $R(t)$ in (\ref{transforme}). 

The complex matter supermultiplets $Z^A (t, \eta, \bar\eta)$ and 
$\bar Z^{\bar A}(t, \eta, \bar \eta) = (Z^A)^\dagger$ consist of a set of 
spatially homogeneous matter fields $z^A(t)$ and $\bar z^{\bar A}(t)$ 
(A= 1,2,..,n), four fermionic degrees of freedom $\chi^A(t)$, 
$\bar\chi^{\bar A}(t)$, $\phi^A(t)$ and $\bar \phi^{\bar A}(t)$, as well as 
bosonic auxiliary fields
$F^A(t)$ and $\bar F^{\bar A}(t)$.

The components of the matter superfields $ Z^A(t, \eta, \bar\eta) $ and 
$\bar Z^{\bar A}(t, \eta, \bar\eta)$ may be written as
\begin{equation}
 Z^A = z^A(t) + i\eta \chi^{\prime A}(t)+ i\bar\eta\phi^{\prime A}(t) + 
F^{\prime A}(t)
\eta\bar\eta,
\end{equation} 

\begin{equation}
 \bar Z^{\bar A}= \bar z^{\bar A}(t) + i\eta \bar\phi^{\prime \bar A}(t) + 
i\bar\eta \bar\chi^{\prime \bar A}(t) + \bar F^{\prime \bar A}(t)
\eta \bar\eta, 
\end{equation}
where
\begin{eqnarray}
&& \chi^{\prime A}(t) = N^{1/2} R^{-3/2} \chi^A(t)\, ,\qquad 
\phi^{\prime A}(t)= N^{1/2} R^{-3/2} \phi^A(t) \, , \nonumber \\
&& F^{\prime A}(t)= NF^A - \frac{1}{2 \sqrt R^3}
(\psi \chi^A - \bar\psi \phi^A)  \, .\nonumber 
\end{eqnarray}
The transformation rule for the superfields $Z^A(t, \eta, \bar\eta)$ and
$\bar Z^{\bar A}(t, \eta, \bar\eta)$ may be written as  
\begin{eqnarray}  
\delta Z^A &=& {I\!\!L} \dot Z^A + \frac{i}{2} D_{\bar\eta} {I\!\!L} D_\eta
Z^A + \frac{i}{2}D_\eta {I\!\!L} D_{\bar\eta} Z^A \, , \label{zeta}\\
\delta \bar Z^{\bar A} &=& {I\!\!L} \dot {\bar Z}^{\bar A} + 
\frac{i}{2} D_\eta {I\!\!L}D_{\bar\eta} \bar Z^{\bar A} + 
\frac{i}{2} D_{\bar\eta} {I\!\!L} D_\eta \bar Z^{\bar A}\, . \label{barzeta}
\end{eqnarray} 

 The most general interacting action for the set of complex homogeneous 
scalar supermultiplets $Z^A(t, \eta, \bar \eta)$ with the scale superfactor
${I\!\!R}(t, \eta, \bar \eta)$ up to the second-order derivatives has the 
following form  

\begin{eqnarray}
S &=& \int \left \{ \Phi \left [  
{I\!\!N}^{-1} {I\!\!R} 
D_{\bar\eta} {I\!\!R} D_\eta {I\!\!R}  - \sqrt{k} {I\!\!R}^2 \right. \right.
\nonumber\\
&& \mbox{} \left.  \left. - \frac{1}{2} \left \{  D_{\bar\eta} \left( 
{I\!\!N}^{-1} {I\!\!R}^2  D_\eta {I\!\!R} \right) -  D_\eta 
\left( {I\!\!N}^{-1} {I\!\!R}^2 D_{\bar\eta} {I\!\!R} \right) \right \} 
\right ]  \right. \nonumber\\ 
 & & \mbox{}\left.  + \frac{1}{2\alpha} {I\!\!N}^{-1} {I\!\!R}^3 
\left [ D_{\bar\eta} 
\bar Z^{\bar A} D_\eta Z^B + D_{\bar\eta} Z^B D_\eta \bar Z^{\bar A} 
\right ] \left[- \Phi^{-1} \frac{\partial \Phi}{ \partial \bar Z^{\bar A}} 
  \frac{\partial \Phi}{\partial Z^B} +
\frac{\partial^2 \Phi}{\partial \bar Z^{\bar A} \partial Z^B }  \right] 
\right. \nonumber \\  
&& \mbox{} \left.
- \frac{2 {I\!\!R}^3}{\kappa^3} | g(Z) |^{\alpha} 
 + \frac{1}{4}{I\!\!N}^{-1} {I\!\!R}^3 \Phi^{-1}D_{\bar\eta} 
\Phi D_{\eta} \Phi \right \} d \eta d \bar \eta dt \, . \label{action-s} 
\end{eqnarray}
This interaction depends on two arbitrary superfunctions 
$\Phi(Z^A, \bar Z^{\bar A})$, $g(Z^A)$ which is dimensionless 
superpotential and of the arbitrary parameter $\alpha$. 
The action (\ref{action-s}) is invariant under the local conformal 
supersymmetric transformations (\ref{superfunction}). If we make the
following Weyl redefinitions
\begin{eqnarray}
&&{I\!\!N} \to  \exp (\frac{\alpha {I\!\!K}}{6}) {I\!\!N}\, , \qquad
{I\!\!R} \to  \exp (\frac{\alpha {I\!\!K}}{6}) {I\!\!R}, 
\label{weyl}\\
&&\Phi \exp (\frac{\alpha {I\!\!K}}{3}) = -\frac{3}{\kappa^2}, \nonumber
\end{eqnarray}
then, our superfield action (\ref{action-s}) takes the form

\begin{eqnarray}
S &=& \int \left \{ -\frac{3}{\kappa^2} {I\!\!N}^{-1} {I\!\!R} D_{\bar\eta}
{I\!\!R} D_{\eta} {I\!\!R} + \frac{3}{\kappa^2} \sqrt{k} {I\!\!R}^2 - 
\frac{2}{\kappa^3}{I\!\!R}^3 e^{\frac{\alpha G}{2}} \right.\nonumber\\
&& +\left. \frac{1}{2\kappa^2}{I\!\!N}^{-1}{I\!\!R}^3 
G_{\bar A B}\left [ D_{\bar\eta} \bar Z^{\bar A}
D_{\eta} Z^B + D_{\bar\eta} Z^B D_{\eta} \bar Z^{\bar Z}\right ]  
\right \} d \eta 
d \bar\eta dt \, .
\label{super-action}
\end{eqnarray}
In this action the total derivative no appear because is not necessary in our
study.
The action (\ref{super-action}) is determined only by terms of one arbitrary 
K\"ahler 
superfunction $G(Z^A, \bar Z^{\bar A})$, and it is a special
combination of ${I\!\!K}(Z^A, \bar Z^{\bar A})$ and $g(Z^A)$, where
\begin{equation}
G(Z, \bar Z) = {I\!\!K}(Z, \bar Z) + log|g(Z)|^2
\end{equation}
is invariant under the transformations
\begin{eqnarray}
g(Z) &\to& g \exp{f(Z)}, \nonumber\\
{I\!\!K}(Z, \bar Z) &\to& {I\!\!K}(Z, \bar Z) - f(Z) - \bar f(\bar Z)
\end{eqnarray}
with the K\"ahler potential ${I\!\!K}(Z,\bar Z)$ defined on the complex
superfield $Z^A$ related to $\Phi(Z, \bar Z)$ (\ref{weyl}). The superfunction 
$G(Z, \bar Z)$ and their transformations are the generalizations of the 
K\"ahler function $G(z, \bar z) = K(z, \bar z) + log|g(z)|^2$ defined on the 
complex manifold and has the Taylor expansion respect to $\eta, \bar\eta$ 
\begin{eqnarray}
G(Z, \bar Z)&=& G(z,\bar z) + G(z, \bar z)_A(Z^A - z^A) + 
G(z, \bar z)_{\bar A}(\bar Z^{\bar A} - \bar z^{\bar A}) \nonumber\\
& &+ \frac{1}{2} G(z, \bar z)_{AB} (Z^A - z^A)(Z^B - z^B)+
\frac{1}{2} G(z, \bar z)_{\bar A \bar B} (\bar Z^{\bar A} - \bar z^{\bar A})
(\bar Z^{\bar B} - \bar z^{\bar B}) \nonumber\\
 &&+ G(z, \bar z)_{\bar A B} 
(\bar Z^{\bar A} - \bar z^{\bar A})(Z^B - z^B), 
\end{eqnarray}
where the first term in the expansion is the K\"ahler function.
Derivatives of the K\"ahler function are denoted by 
$\frac{\partial G}{\partial z^A} = G,_A \equiv G_A$, 
$\frac{\partial G}{\partial {\bar z^{\bar A}}} 
= G,_{\bar A} \equiv G_{\bar A}$, 
$\frac{\partial^n G}{\partial z^A \partial z^B \partial {\bar z^{\bar C}}
 \cdots \partial {\bar z^{\bar D}}} = G,_{AB \bar C \cdots \bar D} 
\equiv G_{AB \bar C \cdots \bar D}$ and the 
K\"ahler metric is $G_{A \bar B}$ = $G_{\bar B A}$ = $K_{A \bar B}$, 
the inverse K\"ahler metric $G^{A \bar B}$, such as 
$G^{A \bar B}G_{\bar B D} = \delta^A_B$, can be used to define 
$G^A \equiv G^{A \bar B} \, G_{\bar B}$ and $G^{\bar B} \equiv 
G_A G^{A \bar B}$.

Perhaps, it is important to mention, that in supergravity it is also possible 
to introduce Weyl transformations $\phi(z, \bar z) = -\frac{3}{\kappa^2} 
\exp({- \frac{\alpha}{3} K(z, \bar z)})$ and the vierbein $e^a_\mu \to 
\exp{\frac{\alpha}{6}K(z, \bar z)}$ with an arbitrary parameter $\alpha$
(see eq.\ref{weyl}).
However, the terms in the supergravity action can not be represented 
by the K\"ahler function $G(z, \bar z)$, which is due to the scalar
curvature term, the kinetic term in the complex fields and auxiliary 
fields $A_\mu$ in the supergravity multiplet are eliminate only if$^8$ 
$\alpha = 1$. 
The action (\ref{super-action}) is invariant under the local conformal 
transformations (\ref{superfunction}) if the superfields are transformed as 
(\ref{coordinate},\ref{genera},\ref{zeta},\ref{barzeta}). After the 
integration over the 
Grassmann variables $\eta, \bar\eta$ the action (\ref{super-action}) becomes a
component action with the auxiliary fields $B(t)$, $F^A(t)$ and 
$\bar F^{\bar A}(t)$. These ones may be determined
from the component action by taking the variation with respect to them.
The equations for these fields are algebraical and have the solutions
\begin{eqnarray}
 B&=& -\frac{\kappa}{18 R^2} \bar\lambda \lambda + \frac{\sqrt k}{\kappa} +
\frac{1}{4\kappa R^2} G_{\bar A B}(\bar\chi^{\bar A} \chi^B + 
\phi^B \bar\phi^{\bar A}) - \frac{R}{\kappa^2} e^{\frac{\alpha G}{2}},
\label{auxiliaryB} \\     
 F^D &=& -\frac{\kappa}{2 R^3} (\bar\lambda \phi^D - \lambda \chi^D) -
\frac{1}{R^3} G^{D \bar A} G_{\bar A B C} \chi^C \phi^B + 
\frac{2}{\kappa} G^{D \bar A} (e^{\frac{\alpha G}{2}}),_{\bar A} \, .
\label{auxiliaryF}
\end{eqnarray}
 
After substituting them again into the component action we get the 
following action

\begin{eqnarray}
S &=& \int \left \{ - \frac{3}{\kappa^2} \frac{R (DR)^2}{N} 
- N\, R^3 U(R,z,\bar z) +\frac{2i}{3} \bar\lambda D\lambda
+ \frac{N \sqrt k}{3R} \bar\lambda \lambda 
-\frac{1}{\kappa} e^{\frac{\alpha G}{2}} \bar\lambda \lambda  
  \right. \nonumber\\
& &+ \frac{\sqrt k}{\kappa}\sqrt{R} \left ( \bar\psi \lambda  
- \psi \bar\lambda \right)  
+ \left. \frac{R^3}{N\kappa^2}G_{\bar A B}
D\bar z^{\bar A} Dz^B  + \frac{i}{2\kappa} Dz^B \left(\bar\lambda G_{\bar A B}
\bar\chi^{\bar A} + \lambda G_{\bar A B} \bar\phi^{\bar A}\right ) 
\right. \nonumber\\
& &+ \left. \frac{i}{2\kappa} D\bar z^{\bar A} \left( \bar\lambda G_{\bar A B}
 \phi^B
+ \lambda G_{\bar A B} \chi^B \right) - \frac{i}{\kappa^2} G_{\bar A B} 
\left( \bar\chi^{\bar A} \tilde D \chi^B + \bar\phi^{\bar A} \tilde D \phi^B 
\right ) \right. \nonumber\\
& &- \left. \frac{N}{\kappa^2 R^3} 
R_{\bar A B \bar C D} \bar\chi^{\bar A} \chi^B \bar\phi^{\bar C} \phi^D 
 - \frac{i}{4\kappa \sqrt{R^3}} \left( \psi \bar \lambda - 
\bar \psi \lambda \right ) G_{\bar AB} \left( \bar \chi^{\bar A} \chi^B + \phi^B 
\bar\phi^{\bar A} \right) \right.
\nonumber\\
& &+ \left. \frac{3N}{16\kappa^2 R^3} \lbrack G_{\bar A B} \left 
(\bar\chi^{\bar A} 
\chi^B + \phi^B \bar\phi^{\bar A}\right) \rbrack^2 
+ \frac{3 \sqrt k}{2 \kappa^2 R}G_{\bar A B}
\left (\bar\chi^{\bar A} \chi^B + \phi^B \bar\phi^{\bar A}\right )
\right. \nonumber\\
& &- \left.  \frac{3N}{2\kappa^3}
e^{\frac{\alpha G}{2}} G_{\bar A B} \left( \bar\chi^{\bar A} \chi^B + 
\phi^B \bar\phi^{\bar A}\right )
- \frac{2N}{\kappa^3} (e^{\frac{\alpha G}{2}}),_{AB}
\chi^A \phi^B 
 - \frac{2}{\kappa^3} N (e^{\frac{\alpha G}{2}}),_{\bar A \bar B}
\bar\phi^{\bar A} \bar\chi^{\bar B} \right. \nonumber\\
& &- \left.  \frac{2}{\kappa^3}N (e^{\frac{\alpha G}{2}}),_{\bar A B}
\left ( \bar\chi^{\bar A} \chi^B + \phi^B \bar\phi^{\bar A}\right ) 
- \frac{N}{\kappa^2}  \bar\lambda \lbrack (e^{\frac{\alpha G}{2}}),_A \phi^A 
+ (e^{\frac{\alpha G}{2}}),_{\bar A}\bar\chi^{\bar A}\rbrack \right.\nonumber\\
& &+ \left. \frac{N}{\kappa^2} \lambda \lbrack (e^{\frac{\alpha G}{2}}),_A 
\chi^A + (e^{\frac{\alpha G}{2}}),_{\bar A} \bar\phi^{\bar A}\rbrack
- \frac{\sqrt{R^3}}{\kappa^2} \left( \bar\psi \lambda - \psi \bar\lambda
\right ) e^{\frac{\alpha G}{2}} \right. \nonumber\\
&&+ \left.  \frac{\sqrt{R^3}}{\kappa^3}(e^{\frac{\alpha G}{2}}),_A
\left ( \psi \chi^A - \bar\psi \phi^A \right ) 
 + \frac{\sqrt{R^3}}{\kappa^3} (e^{\frac{\alpha G}{2}}),_{\bar A} 
\left (\psi \bar \phi^{\bar A} - \bar \psi \bar \chi^{\bar A}\right ) 
\right \} dt, 
\label{action-good} 
\end{eqnarray}
where $DR = \dot R - \frac{i\kappa}{6\sqrt R}\left ( \bar\psi \lambda + 
\psi\bar \lambda\right )$, $Dz^A = \dot{z}^A - \frac{i}{2\sqrt{R^3}}
\left(\bar\psi \phi^A + \psi \chi^A\right )$, 
$D\chi^B= \dot {\chi}^B - \frac{i}{2}V \chi^B$, $D\phi^B = 
\dot {\phi}^B + \frac{i}{2}V \phi^B$, 
$D\lambda = \dot \lambda + \frac{i}{2}V \lambda$,
$\tilde D \chi^B = D \chi^B + \Gamma^B_{CD} \dot {z}^C \chi^D$,
$\tilde D \phi^B = D\phi^B + \Gamma^B_{CD} \dot {z}^C \phi^D$,
$R_{\bar A B \bar C D}$ is the curvature tensor of the K\"ahler manifold 
defined by 
the coordinates $z^A, \bar z^{\bar B}$ with the metric $G_{A \bar B}$, and
$\Gamma^B_{CD}=G^{B \bar A}\, G_{\bar ACD}$ are the Christoffel symbols 
in the definition on covariant derivative and their complex conjugate. 
Besides, the potential term $U(R,z, \bar z )$ read as
\begin{equation}
 U(R,z, \bar z) = -\frac{3 k}{\kappa^2 R^2} + \frac{6\sqrt k}{\kappa^3 R} 
e^{\frac{\alpha G}{2}} + V_{eff}(z, \bar z), 
\label{superpotential}
\end{equation}
where the effective potential of the scalar matter fields is
\begin{equation}
V_{eff} = \frac{4}{\kappa^4} \lbrack 
(e^{\frac{\alpha G}{2}}),_{\bar A} 
G^{\bar A D} (e^{\frac{\alpha G}{2}}),_D - \frac{3}{4}e^{\alpha G} \rbrack = 
\frac{e^{\alpha G}}{\kappa^4} \lbrack \alpha^2 G^A G_A - 3\rbrack \, .
\label{effective-potential}
\end{equation}
In the action (\ref{action-good}) the K\"ahler function is a function of 
scalar fields, i.e $G(z, \bar z)$. 
When the parameter $\alpha = 1$ we obtain the effective potential for 
supergravity coupling with scalar matter fields$^8$. From 
(\ref{superpotential}) we can see, that when 
$k = 0$, $U(R,z, \bar z) = V_{eff}(z, \bar z)$.

In order to discuss the implications of spontaneous supersymmetry breaking
we need to display the superpotential (\ref{superpotential}) in terms of
the auxiliary fields

\begin{equation}
{\cal U}(R,z,\bar z)= \frac{\bar F^{\bar A} G_{\bar A B} F^B}{\kappa^2} 
- \frac{3 B^2}{R^2} \, ,
\label{eff}
\end{equation}
where the bosonic terms (\ref{auxiliaryB}-\ref{auxiliaryF}) 
are read now as
\begin{eqnarray}
B&=& \frac{\sqrt k}{\kappa} 
- \frac{R}{\kappa^2} e^{\frac{\alpha G}{2}}, \label{auxiliary1}\\
F^A&=& \frac{\alpha}{\kappa} e^{\frac{\alpha G}{2}}\, G^A \, .
\label{auxiliary2}
\end{eqnarray}
The supersymmetry is a spontaneous breaking, if the auxiliary fields 
(\ref{auxiliary2}) of the matter
supermultipletes get non-vanishing vacuum expectation values. According to 
our assumption at the minimum in (\ref{eff}) 
${\cal U}(R,<z^A>,<\bar z^{\bar A}>)=0,$ but $<F^A>\not=0$ and, thus, 
the supersymmetry is broken. The measure of this breakdown is the term 
$(- \frac{1}{\kappa} e^{\frac{\alpha}{2}G(<z^A>, <\bar z^{\bar A}>)})
\bar \lambda \lambda $ in the action (\ref{action-good}). Besides, we can 
identify
\begin{equation}
m_{3/2} = \frac{1}{\kappa} e^{\frac{\alpha}{2}G(z^A, \bar z^{\bar A})}
\end{equation}
as the gravitino mass in the effective supergravity theory$^8$.  

Hence, we can see, that in our model the conformal time supersymmetry 
(\ref{superfunction}), being subgroup of space-time SUSY, gives a general 
mechanism of spontaneous breaking of this SUSY$^8$. If we consider 
dilatonic fields in K\"ahler function, then the parameter $\alpha$ may be 
viewed as a dilaton coupling constant$^9$. 

In the future it will be shown, that in the quantum level, the 
supersymmetric quantum mechanics action (\ref{action-good}) describes the 
vacuum states of supergravity, as well as supersymmetric cosmological FRW 
model.

{\Large Acknowledgments}

We are grateful to D.V. Shirkov, E.A. Ivanov, O. Obreg\'on, S. Krivanos,
A.I. Pashnev, I.C. Lyanzuridi and M.P. Ryan for their interest in this paper. 
This work was partially supported by  CONACyT, grant No. 3898P--E9608.

\newpage
{\Large References}
\begin{enumerate}
\item{} B.S. De Witt, {\it Phys. Rev.} {\bf 160}, 1143 (1967); J. A. Wheeler,
	in {\it Relativity Groups and Topology}, edited by C. De Witt and B.
	De Witt (Gordon and Breach, New York, 1969); M.P. Ryan, 
	{\it Hamiltonian Cosmology} (Springer Verlag, Berlin, 1972).
\item{}  A. Mac\'{\i}as, O. Obreg\'on and M.P. Ryan Jr,  {\it Class. 
             Quantum Grav.} {\bf 4}, 1477 (1987); P.D. D'Eath  
	     and D.I. Hughes, {\it Phys. Lett. } {\bf B214}, 498 (1988). 
\item{} R. Graham,  {\it Phys. Rev. Lett.} {\bf 67}, 1381 (1991).
\item{}P.D. D'Eath,  {\it Supersymmetric Quantum Cosmology} 
              (Cambridge: Cambridge University Press, 1996).
\item{}P.V. Moniz, {\it supersymmetric quantum Cosmology} {\it Int.
               J. of Mod. Phys. } {\bf A11}, 4321-4382 (1996).
\item{} O. Obreg\'on, J.J. Rosales and V.I. Tkach, {\it Phys. Rev. } 
              {\bf D53}, R1750 (1996); V.I. Tkach, J.J. Rosales and 
	      O. Obreg\'on, {\it Class. Quantum Grav.} {\bf 13}, 2349 (1996).
\item{}V.I. Tkach, O. Obreg\'on and J.J. Rosales, {\it Class. Quantum. Grav.}
	{\bf 14}, 339 (1997).
\item{}E. Cremmer, B. Julia, J. Scherk, S. Ferrara, L. Girardello and
               P. van Niewenhuizen,  {\it Nucl. Phys. } {\bf B147}, 105 (1979);
	       V.S. Kaplunovsky, J. Louis, {\it Phys. Lett } {\bf B306}, 
	       269 (1993);
	       P. Nath, R. Arnowitt and C. Chamseddine  {\it Applied N=1
                Supergravity} World scientific Publishing, 1984.
\item{}D. Garfinkle, G. Horowitz, A. Strominger, { \it Phys. Rev. }
	{\bf D43}, 3140 (1991); A. Mac\'{\i}as and T. Matos, 
	{\it Class. Quantum
	Grav.} {\bf 13}, 345 (1996); V. Halyo and E. Halyo, {\it Phys. Lett. 
	} {\bf B382}, 89 (1996).
\end{enumerate}
\end{document}